\pdfoutput=1
\documentclass[
aps,preprintnumbers,amsmath,amssymb,prd,nofootinbib,reprint]{revtex4-1}

\usepackage{graphicx}
\usepackage{dcolumn}

\usepackage[colorlinks=true,allcolors=blue]{hyperref}

\DeclareMathOperator{\Tr}{Tr}
\usepackage{slashed}

\begin{document}


\graphicspath{{./images/}}
\def\a{\alpha}
\def\b{\beta}

\newcommand{\gsim}{ \mathop{}_{\textstyle \sim}^{\textstyle >} }
\newcommand{\lsim}{ \mathop{}_{\textstyle \sim}^{\textstyle <} }
\newcommand{\vev}[1]{ \left\langle {#1} \right\rangle }
\newcommand{\bra}[1]{ \langle {#1} | }
\newcommand{\ket}[1]{ | {#1} \rangle }
\newcommand{\et}[1]{  {#1} \rangle }
\def\diag{\mathop{\rm diag}\nolimits}
\def\SU{\mathop{\rm SU}}
\def\U{\mathrm{U}}
\def\tr{\mathop{\rm tr}}

\def\CA{{\cal A}}
\def\CC{{\cal C}}
\def\CD{{\cal D}}
\def\CE{{\cal E}}
\def\CF{{\cal F}}
\def\CG{{\cal G}}
\def\CH{{\cal H}}
\def\CI{{\cal I}}
\def\CK{{\cal K}}
\def\CL{{\cal L}}
\def\CR{{\cal R}}
\def\CM{{\cal M}}
\def\CN{{\cal N}}
\def\CO{{\cal O}}
\def\CQ{{\cal Q}}
\def\CS{{\cal S}}
\def\CT{{\cal T}}
\def\cU{\mathcal{U}}
\def\cV{\mathcal{V}}

\newcommand{\bH}{\mathbb{H}}
\newcommand{\bC}{\mathbb{C}}
\newcommand{\bR}{\mathbb{R}}
\newcommand{\bZ}{\mathbb{Z}}

\def\SU{\mathrm{SU}}
\def\SO{\mathrm{SO}}
\def\tr{\mathop{\mathrm{tr}}\nolimits}
\def\Tr{\mathop{\mathrm{Tr}}\nolimits}
\def\II{I\hspace{-.1em}I}
\def\vev#1{\langle#1\rangle}
\def\cH{{\cal H}}
\def\ff{\mathfrak{f}}
\def\fg{\mathfrak{g}}
\def\CP{\mathbb{CP}}

\def\bT{\mathbb{T}}

\def\su{\mathfrak{su}}
\def\u{\mathfrak{u}}

\newcommand{\MY}[1]{\textbf{\color{green} [{\sc MY}: {#1}]}}
\newcommand{\KY}[1]{\textbf{\color{red} [{\sc KY}: {#1}]}}

\def\beq#1\eeq{\begin{align}#1\end{align}}
\def\alert#1{{\color{red}[#1]}}

\newcommand{\Kahler}{K\"ahler }

\newcommand{\modA}{\mathbb{E}}

\newcommand{\khalf}{k\to k+1}

\newcommand{\el}{\mathsf{e}}

\preprint{
IPMU 17-0082; 
}

\title{
Anomaly constraints on deconfinement and chiral phase transition}

\author{Hiroyuki Shimizu} 
\author{Kazuya Yonekura}
\affiliation{Kavli IPMU (WPI), UTIAS, 
The University of Tokyo, 
Kashiwa, Chiba 277-8583, Japan}


\begin{abstract} 
We study constraints on thermal phase transitions of ${\rm SU}(N_c)$ gauge theories by using 
the 't~Hooft anomaly involving the center symmetry and chiral symmetry.
We consider two cases of massless fermions: 
(i)  adjoint fermions, and (ii) $N_f$ flavors of fundamental fermions with 
a nontrivial greatest common divisor ${\rm gcd}(N_c,N_f) \neq 1$.
For the first case (i), we show that the chiral symmetry restoration in terms of the standard Landau-Ginzburg effective action 
is impossible at a temperature lower than that of deconfinement. 
For the second case (ii), we introduce a modified version of the center symmetry which we call center-flavor symmetry, and draw
similar conclusions under a certain definition of confinement. 
Moreover, at zero temperature, our results give a partial explanation of the appearance of dual magnetic gauge group
in (supersymmetric) QCD when ${\rm gcd}(N_c,N_f) \neq 1$.

\end{abstract}

\maketitle


\section{Introduction and summary}
Thermal phase transition in gauge theories is a very interesting and important subject.
Theoretically, it is related to the mystery of how strong dynamics works in confinement and chiral symmetry breaking.
Phenomenologically, the nature of phase transition affects cosmological observables
such as dark matter abundance. It might even provide the dark matter itself via QCD effects~\cite{Witten:1984rs}.

The standard way to study chiral symmetry restoration is as follows~\cite{Pisarski:1983ms}. 
The quark bilinear $\Phi \sim \psi\psi$, where $\psi$ represents left-handed fermions, is believed to be the most relevant order parameter for the chiral symmetry breaking.
This operator $\Phi$ is treated as the effective degrees of freedom near the critical temperature $T_{\text{chiral}}$, 
and the phase transition is described by a Landau-Ginzburg effective Lagrangian 
\beq
{\cal L}_{\rm eff} &= \tr (\partial_i \Phi^\dagger \partial_i \Phi) +V(\Phi) \nonumber \\
V(\Phi) &= c_{0}\tr ( \Phi^\dagger  \Phi) +c_1 [\tr (\Phi^\dagger \Phi)^2]+
c_2[\tr (\Phi^\dagger \Phi) ]^2    \nonumber \\
&+c_{\rm anom}[\det(\Phi)]^{t_R}+\cdots, \label{eq:LG}
\eeq
where the coefficients depend on temperature $T$ and in particular $c_{0} \propto (T-T_{\text{chiral}})$, and $t_R$ is the Dynkin index of of the left-handed fermion representation.
However, because of the strong coupling, it is not easy to see whether
such a scenario is likely or not. It is conceivable that deconfinement happens at a lower temperature.
If so, we lose intuitive reasons for treating the composite $\Phi$ as the effective elementary degrees of freedom. 
These questions may be rigorously asked in theories where center symmetry is well-defined, such as adjoint QCD, 
QCD with imaginary baryon chemical potential~\cite{Roberge:1986mm}, or QCD in the large $N_c$ limit.

Recently, a very remarkable paper~\cite{Gaiotto:2017yup} appeared which studied
phase transitions in pure Yang-Mills. They used a mixed 't~Hooft anomaly 
between the ${\rm CP}$ symmetry and the 1-form center symmetry to constrain possible phase transitions at
the theta angle $\theta=\pi$. Under reasonable assumptions about the dynamics of pure Yang-Mills, 
the ${\rm CP}$ symmetry cannot be restored below the temperature at which the center symmetry
is broken, i.e., deconfinement. See the original paper for more careful discussions.

In this paper, we point out that a similar discussion is possible for chiral symmetry.
The mixed 't~Hooft anomaly between chiral and center symmetry is also known~\cite{Gaiotto:2014kfa},
so we can repeat the argument of \cite{Gaiotto:2017yup} when the center symmetry exists,
such as $\SU(N_c)$ gauge theories with $n_f$ fermions in the adjoint representation.
We will see that chiral symmetry restoration by \eqref{eq:LG} cannot happen below the deconfinement temperature. 

When there are $N_f$ fermions in the fundamental representation of $\SU(N_c)$, the center symmetry no longer exists.
However, we argue that there is a more subtle ``symmetry" which mixes the center symmetry and flavor symmetry,
by using the fact that the fermions are in the representation of $[\SU(N_c) \times \SU(N_f)_V]/\bZ_n$ where 
$n:={\rm gcd}(N_c,N_f)$ is the greatest common divisor.
The division by $\bZ_n$ leads to what we call ``center-flavor symmetry".
Then, we get similar constraints as in the case of adjoint fermions, under a 
technical definition of confinement in terms of the quantum fluctuations of the gauge field in confining phase. 
This has implications even at zero temperature. If the chiral symmetry is not broken, we need 
dynamical gauge fields to match the anomaly of the center-flavor symmetry. 
This partially explains the reason why there appears dual magnetic gauge group in Seiberg's description of
supersymmetric QCD~\cite{Seiberg:1994pq,Intriligator:1995au}.
In a sense, we can  directly see the existence of gauge fields via the 't~Hooft anomaly.

\section{$\SU(N_c)$ with adjoint fermions}
In this section, we consider $\SU(N_c)$ gauge theories 
with $n_f$ massless Weyl fermions in the adjoint representation. 

\subsection{'t~Hooft anomaly of chiral and center symmetry}\label{sec:anomadj}
Here we describe the mixed 't~Hooft anomaly of chiral and center symmetry \cite{Gaiotto:2014kfa}.

\paragraph{Chiral symmetry.}
Classically the theory has $\U(n_f)=[\U(1)_A \times \SU(n_f)]/\bZ_{n_f}$ chiral symmetry acting on the fermions. The $\U(1)_A$ is quantum mechanically broken to the anomaly-free subgroup $\bZ^{\rm axial}_{2N_c n_f}$, whose generator acts on fermions $\psi$ via
\begin{equation}
\bZ^{\rm axial}_{2N_c n_f}: \psi \to \exp\biggl{(}\frac{2\pi i}{2N_c n_f} \biggr{)}\psi. \label{eq:axi}
\end{equation}
Thus the chiral symmetry of the theory is reduced to
$
[\SU(n_f) \times \bZ^{\rm axial}_{2N_c n_f}] / \bZ_{n_f} 
$.

The order parameter of the breaking is $\Phi_{ab}=\psi_a \psi_b~(a,b=1,\cdots,n_f)$ which behave as
\begin{equation}
\det \Phi_{ab} = \text{const} \cdot (e^{ i\theta})^{\frac{1}{N_c}}.
\end{equation}
Assuming it is nonzero, there are $N_c$ distinct connected components in the moduli space of vacua.
The generator of $\bZ^{\rm axial}_{2N_c n_f}$ transformation is implemented by the theta angle rotation  
$\theta \to \theta + 2\pi$ which permutes the $N_c$ connected components.

The continuous part $\SU(n_f)$ of the chiral symmetry is also broken by the vacuum expectation values of the matrix $(\Phi_{ab})$, 
which produce Goldstone bosons at each connected component. However, the details of this breaking do not play any role in the following discussion.

\paragraph{Center symmetry.}

Since adjoint fermions transform trivially under the center $\bZ_{N_c} \subset \SU(N_c)$, the theory possesses the $\bZ_{N_c}$ center symmetry. The center symmetry is a typical example of 1-form symmetry \cite{Gaiotto:2014kfa}, which acts on line operators in the present theory.

The 1-form center symmetry can be coupled 
to a 2-form background $B \in H^2(X,\bZ_{N_c})$, where $X$ is spacetime, as follows.
For a topologically nontrivial gauge bundle on a manifold $X$, we first take open covers $\{ U_a \}_{a \in A}$ of $X$ 
such that the bundle is trivialized on each of $U_a$. They are glued by transition functions $g_{ab}$
on $U_a \cap U_b$ which we take to be $N_c \times N_c$ matrices in the fundamental representation.
For $\SU(N_c)$ (as opposed to $\SU(N_c)/\bZ_{N_c}$) bundles, they satisfy the standard consistency condition 
$g_{ab}g_{bc}g_{ca}=1$ if $U_a \cap U_b \cap U_c \neq \varnothing$.
However, when we consider $\SU(N_c)/\bZ_{N_c}$ bundles, we have
\beq
 g_{ab}g_{bc}g_{ca} = \exp\left( \frac{2\pi i w_{abc}^{\rm gauge}}{N_c} \right) ,~~~~~
 w_{abc}^{\rm gauge}  \in \bZ_{N_c}. \label{eq:gluingconsistency}
\eeq
This is allowed because $ \exp(2\pi i w_{abc}^{\rm gauge}/N_c) $ is in the center $\bZ_{N_c} \subset \SU(N_c)$ and hence it is trivial in $\SU(N_c)/\bZ_{N_c}$.
These $w_{abc}^{\rm gauge}$ give an element of cohomology group 
\beq
w_2^{\rm gauge}  \in H^2(X,\bZ_{N_c})
\eeq
which gives the obstruction to uplifting an $\SU(N_c)/\bZ_{N_c}$ bundle to an $\SU(N_c)$ bundle.

Including the background field $B \in H^2(X,\bZ_{N_c})$ for the center 1-form symmetry corresponds to considering 
gauge bundles which satisfy $B=w_2^{\rm gauge} $. Namely, we perform path integral under this topological condition for the gauge field.

\paragraph{Mixed 't~Hooft anomaly.}  
Let us describe the mixed 't~Hooft anomaly between the axial symmetry $\bZ^{\rm axial}_{2N_c n_f}$ and the center symmetry~\cite{Gaiotto:2014kfa}.
Under the axial rotation \eqref{eq:axi}, the standard Fujikawa's argument tells us that the path integral measure $Z(X)$ changes as
\beq
Z(X) \to Z(X) \exp(  2\pi i \int_X \frac{1}{8\pi^2}\tr F \wedge F)
\eeq
where $F$ is the gauge field strength and the trace is in the fundamental representation.

If $B=w_2^{\rm gauge}=0$, the above phase factor is trivial because the instanton number is integral. However, when we turn on the background field $B=w_2^{\rm gauge} \neq 0$, we have (on a manifold like $T^4$ \cite{Witten:2000nv})
$
\frac{1}{8\pi^2}\int_X \tr F \wedge F = - \frac{1}{2N}\int_X B \wedge B \mod 1
$ and hence
\beq
Z(X) \to Z(X) \exp(   - \frac{2\pi i}{2N}\int_X B \wedge B). \label{eq:adjanom}
\eeq
This represents the mixed 't~Hooft anomaly.

\paragraph{Low energy behavior.} 
There is an immediate consequence of the above mixed 't~Hooft anomaly. 
It is impossible that the low energy limit have a trivial gapped vacuum with both the chiral and center symmetries unbroken.

By looking at the two-loop beta function, a likely scenario is as follows \cite{Shifman:2013yca}. When $n_f \leq 2$, the chiral symmetry is spontaneously broken.
When $n_f=5$, it flows to a conformal fixed point. The cases $n_f=3,4$ are unclear, but $n_f=4$ may have a conformal fixed point.

\subsection{Constraints on phase transition}
Next we use the mixed 't~Hooft anomaly of the $\bZ^{\rm axial}_{2N_c n_f}$ and the $\bZ_{N_c}$ center symmetry to constrain the thermal phase transition. We reduce the theory along the thermal circle $S^1_{T}$ and obtain the effective theory on $\bR^3$ .The center symmetry now splits into two global symmetries. 
One is the 0-form center symmetry $\bZ_{N_c}^\text{0-form}$ acting on the Polyakov loop $L= \tr_{{\bf N_c}} {\rm P} \exp(i \int_{S^1_T} A)$. The other is the 1-form center symmetry $\bZ_{N_c}^\text{1-form}$ acting on space-like Wilson loops extending along $\bR^3$.  

The three-dimensional effective theory still has the mixed ``triangle" 't~Hooft anomaly among the three symmetries 
$(\bZ_{N_c}^\text{0-form})(\bZ_{N_c}^\text{1-form})(\bZ^{\rm axial}_{2N_c n_f})$ obtained by dimensional reduction of \eqref{eq:adjanom}.
The anomaly forbids the three symmetries to be simultaneously preserved. 

\paragraph{High/Low temperature phases.} We summarize the symmetry breaking in the high/low temperature limit. First, we note that the fermions have the anti-periodic boundary condition along the thermal circle and have no zero-modes on $\bR^3$. 
Then, when the temperature is sufficiently high, they can be safely integrated out. 
The axial $\bZ^{\rm axial}_{2N_c n_f}$ is unbroken. 

The remaining degrees of freedom consists of the 3d Yang-Mills and the periodic scalars coming from the gauge field in the direction $S^1_T$.  
The scalars get the effective potential at the one-loop level such that the 
$\bZ_{N_c}^\text{0-form}$ is broken. 
The 3d Yang-Mills is expected to confine with the area law for space-like Wilson loops. Therefore, at extremely high temperature,  
the $\bZ_{N_c}^\text{0-form}$ is broken, while the $\bZ_{N_c}^\text{1-form} \times \bZ^{\rm axial}_{2N_c n_f}$ are unbroken.

At very low temperature, the theory can be regarded as four-dimensional. We focus on the case in which the theory confines
and the chiral symmetry is broken.
Then $\bZ^{\rm axial}_{2N_c n_f}$ is broken, while $\bZ_{N_c}^\text{0-form} \times \bZ_{N_c}^\text{1-form}$ is not.

The summary is in the following table.
\begin{table}[htb]
\begin{center}
\begin{tabular}{cccc}
Symmetry & Low $T$ & Intermediate & High $T$ \\
center $\bZ_{N_c}^\text{0-form}$ & unbroken & ? & broken \\
center $\bZ_{N_c}^\text{1-form}$ & unbroken & ? & unbroken \\
axial $\bZ^{\rm axial}_{2N_c n_f}$ & broken & ? & unbroken \\
\end{tabular}
\end{center}
\end{table} 
\paragraph{Inequality for $T_{\text{deconf}}$ and $T_{\text{chiral}}$.} 
We can define at least two critical temperatures: deconfinement temperature 
$T_{\text{deconf}}$ for $\bZ_{N_c}^\text{0-form}$,
and chiral symmetry restoration temperature $T_{\text{chiral}}$ for $\bZ^{\rm axial}_{2N_c n_f}$. 
We don't consider the cases with more than two critical temperatures.

Now, suppose that the chiral symmetry is restored by the Landau-Ginzburg effective action \eqref{eq:LG}.
Then the mixed 't~Hooft anomaly implies the inequality between the two temperatures,
\begin{equation}
T_{\text{deconf}} \leq T_{\text{chiral}}.\label{eq:ineqadj}
\end{equation}
The reason is as follows.
Suppose \eqref{eq:ineqadj} does not hold. Then in the intermediate temperature $T_{\text{chiral}} < T < T_{\text{deconf}}$, both $\bZ_{N_c}^\text{0-form}$ and  $\bZ^{\rm axial}_{2N_c n_f}$ are unbroken. If the physics near $T_{\text{chiral}}$ is described by \eqref{eq:LG}, there is no way to break the 1-form symmetry $\bZ_{N_c}^\text{1-form}$,
and also there is no gapless degrees of freedom in $T_{\text{chiral}} < T < T_{\text{deconf}}$. 
This contradicts with the anomaly because all the symmetries are unbroken and there is no degrees of freedom to match the anomaly.

If \eqref{eq:ineqadj} holds, both $\bZ_{N_c}^\text{0-form}$ and  $\bZ^{\rm axial}_{2N_c n_f}$  are broken in $T_{\text{deconf}} < T < T_{\text{chiral}}$. This is consistent with the mixed 't~Hooft anomaly. But it is a little counter-intuitive to use \eqref{eq:LG} because, intuitively, the gluons and quarks are liberated in 
the deconfining phase $T \sim T_{\text{chiral}} > T_{\text{deconf}} $.

A lattice study \cite{Karsch:1998qj} with $n_f=4$ gave a result consistent with \eqref{eq:ineqadj}. However, that result is not conclusive because the theory with $n_f=4$ may have conformal fixed point~\cite{DelDebbio:2010zz,Shifman:2013yca,DeGrand:2013uha}. Results in a sequence of semiclassical studies of adjoint QCD, 
e.g.~\cite{Kovtun:2007py,Unsal:2007vu,Unsal:2007jx,Poppitz:2009uq,Anber:2011gn,Misumi:2014raa,Anber:2015wha} are consistent with our constraints.

Finally, let us mention two alternative scenarios without assuming \eqref{eq:LG}. They do not require \eqref{eq:ineqadj} .
\begin{enumerate}
\item There is a single first-order phase transition at $T_c=T_{\text{chiral}}=T_{\text{deconf}}$. When we cross the temperature $T_c$, the decofinement transition and the chiral symmetry restoration occurs at the same time.

\item We allow a phase with broken $\bZ_{N_c}^\text{1-form}$ in $T_{\text{chiral}} < T < T_{\text{deconf}}$. 
Namely, we have a Higgs phase for the effective 3d Yang-Mills in the intermediate temperatures as discussed in \cite{Gaiotto:2017yup}. However, this scenario seems difficult in the presence of the order parameter $\Phi_{ab}$.


\end{enumerate}

\section{$\SU(N_c)$ with fundamental fermions}
In this section, we consider $\SU(N_c)$ gauge theories 
with massless fermions in the fundamental and anti-fundamental
representations ${\bf N_c}+\overline{\bf N_c}$. We assume that the flavor number $N_f$ and the color number $N_c$
have a nontrivial greatest common divisor $n:={\rm gcd}(N_f,N_c) \neq 1$ which includes the case $N_c=N_f$,
such as the $\SU(3)$ QCD in the massless limit of up, down and strange quarks.

\subsection{Center-flavor symmetry }
First, we explain a way to introduce non-trivial background fields to detect the anomaly.

When matter fields are in the fundamental representation, it does not make mathematical sense to take transition functions
as in \eqref{eq:gluingconsistency} with $w_{abc}^{\rm gauge} \neq 0$. However, we can avoid this problem by the following trick.\footnote{
The idea similar to here appeared in \cite{Witten:1994cg} where
spinors are put on non-spin 4-manifolds by considering ${\rm spin}^c $ structure.
The gravitational background there corresponds to the flavor background here, and the $\U(1)$ gauge field there corresponds to 
the $\SU(N)$ gauge field here. Analogous interplay between global and gauge symmetries have also appeared 
in recent discussions of topological phases of matter,
see e.g., \cite{Seiberg:2016rsg, Tachikawa:2016xvs,Witten:2016cio}.
Formally, we are going to use the fact that the flavor symmetry acting on gauge invariant operators is $\SU(N_f)_V/\bZ_n$,
and it has the extension $1 \to \SU(N_c) \to [\SU(N_c) \times \SU(N_f)_V]/\bZ_n \to \SU(N_f)_V/\bZ_n \to 1$.
The ideas very close to ours have appeared also in \cite{Benini:2017dus,Komargodski:2017dmc}.
} 
The matter fields are in the bifundamental 
representations ${\bf N_c} \times \overline{\bf N_f}$ of the $\SU(N_c) \times \SU(N_f)_V$ symmetry where $\SU(N_f)_V$ is the diagonal 
subgroup of the $\SU(N_f)_L \times \SU(N_f)_R$ chiral symmetry.
Let
$
n={\rm gcd}(N_c,N_f)
$
be the greatest common divisor of $N_c$ and $N_f$.
There is a subgroup $\bZ_n  \subset \SU(N_c) \times \SU(N_f)_V$ which acts trivially on the fermions.
Then it is possible to consider $[\SU(N_c) \times \SU(N_f)_V]/\bZ_n$ bundles.
(See also \cite{Tanizaki:2017bam} in which $\SU(N_f)_V$ is dynamical.)

More concretely, we consider the following gauge and flavor bundles. The flavor bundle has transition functions $h_{ab}$ satisfying
\beq
 h_{ab}h_{bc}h_{ca} =\exp\left( \frac{2\pi i w_{abc}^{\rm flavor}}{N_f} \right) ,~~~~~
 w_{abc}^{\rm flavor}  \in \bZ_{N_f}.
\eeq
Then we require
\beq
\frac{n}{N_c} w_{abc}^{\rm gauge}=
\frac{n}{N_f}w_{abc}^{\rm flavor} := w_{abc}  \in \bZ_n.
\eeq
Under this condition, the fermions are put on $X$ because the total transition functions $(g \otimes h^\dagger)_{ab}$ satisfy
$(g \otimes h^\dagger)_{ab} \cdot (g \otimes h^\dagger)_{bc} \cdot (g \otimes h^\dagger)_{ca}=1$. 
We call the ``symmetry" corresponding to this background as center-flavor symmetry,
although we do not give Hilbert space interpretation.

\paragraph{Anomaly.}

We can see the existence of the anomaly of center-flavor symmetry by the following concrete setup. 
Compactify the spacetime to 
$X=S^1_T \times S^1_A \times S^1_B \times \bR_C$, where $S^1_T$ will be the temporal direction (i.e., thermal circle) and 
$S^1_A \times S^1_B \times \bR_C$ are the spatial directions. The radii of $S^1_{A,B}$ are taken to be much larger than that of $S^1_T$.

We introduce the flavor background along $S_{T,A,B}$ as follows~\cite{tHooft:1979rtg,Witten:1982df,Witten:2000nv}.
In the direction $S^1_A \times S^1_B$, we introduce the flavor Wilson lines $\Omega_A$ and $\Omega_B$ given as
\beq
\Omega_A &= I_{N_f/n} \otimes \omega_A, \\
\Omega_B &= I_{N_f/n} \otimes \omega_B, 
\eeq
where $I_m$ means the unit $m \times m$ matrix, and 
$\omega_A$ and $\omega_B $ are $n \times n$ matrices with the commutation relation $\omega_A \omega_B =e^{2\pi i/n} \omega_A  \omega_B$. Explicitly,
\beq
&\omega_A =\diag(1, e^{2\pi i /n},e^{4\pi i /n}, \cdots, e^{2(n-1)\pi i /n} ) \\
&\omega_B =(\delta_{i+1,j} )_{1\leq i, j \leq n} 
\eeq
We take the flavor Wilson line in the direction $S^1_T $ to be an imaginary baryonic chemical potential $\mu_B$~\cite{Roberge:1986mm,deForcrand:2002hgr}
\beq
\Omega_T = e^{i \mu_B/N_c} I_{N_f}.
\eeq 
The flavor background is a flat connection. 

For the gauge field, we have a freedom to choose their boundary conditions.
Let $x_C \in \bR_C$ be the coordinate. We impose boundary conditions at $x_C \to \pm \infty$ 
such that the gauge field approaches flat connections represented by gauge Wilson lines $W_T(x_C), W_A(x_C), W_B(x_C)$ as
\beq
W_A(x_C = \pm \infty)  &=  V_A \otimes \omega_A \nonumber \\
W_B(x_C = \pm \infty )  &= V_B \otimes \omega_B \nonumber \\
W_T(x_C = \pm \infty) &= V_{ \pm \infty} \otimes I_n
\eeq
where $\omega_A$ and $\omega_B$ are the same ones as in the flavor background introduced above, and
$(V_A , V_B, V_{\pm \infty})$ are $N_c/n \times N_c/n$ unitary matrices such that $\det W_A = \det W_B=\det W_T=1$, and 
\beq
\det (V_{ \pm \infty}) = \exp( 2\pi i m_{ \pm \infty} /n) \label{eq:defm}
\eeq
for $m_{ \pm \infty} \in \bZ_n$. The $V_A$, $V_B$ and $V_{\pm \infty}$ commute with each other
so that the gauge field is flat at infinity. 

The above configurations give a nontrivial $w_2$ as
\beq
\frac{n}{N_f}\int_{S^1_A \times S^1_B} w^{\rm flavor}_2 = \frac{n}{N_c} \int_{S^1_A \times S^1_B} w^{\rm gauge}_2 =1 \mod n. \nonumber
\eeq
Due to this $w_2$, there are fractional instantons in $X$ with instanton charge 
$(m_{+\infty} - m_{-\infty})/n \mod 1$ \cite{Yamazaki:2017ulc} (see also \cite{Ohmori:2015pua}). 
We remark that explicit instanton solutions are not at all necessary for our discussion, and only the topological data are important.
Then, Atiyah-Patodi-Singer (APS) index theorem states (for generic $V_A,V_B,V_{\pm \infty},\mu_B$)
that there are fermion zero modes such that under the $\bZ^{\rm axial}_{2N_f}$ axial rotation 
\beq
\bZ^{\rm axial}_{2N_f}:~~\psi \to e^{\frac{2 \pi i}{ 2 N_f } } \psi. \label{eq:Arotate}
\eeq
the path integral measure $Z(X)$ gets a phase factor
\beq
Z(X) \to  Z(X)  \cdot \exp\left(2\pi i  (m_{+\infty} - m_{-\infty})/n  \right). \label{eq:keyanomaly}
\eeq
This is the key anomaly for our purposes.

\subsection{Constraints by anomaly}
We would like to discuss some consequences of the anomaly \eqref{eq:keyanomaly}.
\paragraph{Thermal phase transition.}
Because of the anomaly \eqref{eq:keyanomaly}, 
there are constraints on phase transitions. 
First, let us discuss the case of a specific value of $\mu_B$. The fundamental fermions are coupled to the total Wilson line
$W_T \otimes \Omega_T^\dagger$. Now, the effect of center symmetry action $W_T \to e^{-2\pi i /N_c} W_T$ can be compensated by
the shift $\mu_B \to \mu_B-2\pi $. By combining parity in the $S_T$ direction $\mu_B \to -\mu_B$, we find that
there is a $\bZ_2$ symmetry if $\mu_B= \pi$~\cite{Roberge:1986mm,deForcrand:2002hgr}.
This $\bZ_2$ acts on the Polyakov loop $L = \tr_{\bf N_c} W_T$ as $L \to (e^{-2\pi i /N_c} L)^*$ where $ e^{-2\pi i /N_c}$
comes from the center symmetry action and the complex conjugate comes from the parity flip on $S^1_T$.
Thus this is a symmetry whose order parameter is the Polyakov loop, and as we discuss below,
it is broken at high temperature while it is unbroken at low temperature. Therefore, this $\bZ_2$ can be used
for a rigorous definition of deconfinement/confinement phases at $\mu_B=\pi$, 
just as the $\bZ^{ \text{0-form}}_{N_c}$ symmetry of the adjoint fermion theory.

At high temperature, the $\bZ_2$ is spontaneously broken just by standard perturbative computation at finite temperature,
and the minima of the effective potential are at $W_T =I_{N_c} $ and $ e^{2\pi i/N_c} I_{N_c}$
which are related by $\bZ_2$.
Now let us take the boundary conditions as $m_{-\infty}=0$ and $m_{+\infty}=1$. 
These two values are in the two vacua related by the spontaneously broken $\bZ_2$.
Then $m_{-\infty}=0$ and $m_{+\infty}=1$ means that the gauge configuration approaches to these two vacua at $x_C \to \pm \infty$.
The domain wall interpolating the two vacua is the fractional instanton.
 
At very low temperature, we can see that $\bZ_2$ is unbroken as follows. If it were broken,
then by changing the value of $\mu_B$ from $\pi - \epsilon$ to $\pi+\epsilon$ for an infinitesimal $\epsilon$, 
there would be a phase transition from one phase to another which are related by $\bZ_2$.
However, the $\mu_B$ is coupled to the Baryon number and all the particles having nonzero baryon numbers are heavy.
Therefore, the change of $\mu_B$ from $ \pi - \epsilon $ to $\pi + \epsilon$ cannot change the dynamics of low energy physics 
(i.e., the effective theory of pions)
if the temperature is significantly lower than the lowest baryon mass, and hence there are no phase transitions 
associated to the assumed spontaneously broken $\bZ_2$ symmetry. Therefore, the $\bZ_2$ should not be broken.
This consideration is in agreement with the numerical results in \cite{deForcrand:2002hgr} (see Figure~3 of that paper).

The fact that $\bZ_2$ is unbroken at low temperatures means that $W_T$ is well-fluctuating and the vacuum state has overlap with any value of $m_{\pm \infty}$.
 So the boundary conditions $m_{-\infty}=0$ and $m_{+\infty}=1$ become irrelevant at low temperature. 
 
Therefore we have the following situation;
\beq
\begin{array}{cccc}
& \bZ_{2N_f}^{\rm axial} & \bZ_2 & \text{Source of anomaly} \\
\text{Low } T & \text{broken} & \text{unbroken} &  \bZ_{2N_f}^{\rm axial} \text{ breaking} \\
\text{Intermediate} & ? & ? & ? \\ 
\text{High } T & \text{unbroken} & \text{broken} & \text{fractional instanton}
\end{array} \nonumber 
\eeq
Comparison with the case of adjoint fermions is that $\bZ_{2 N_c n_f}^{\rm axial} \leftrightarrow \bZ_{2N_f}^{\rm axial}$, $\bZ_{N_c}^\text{0-form} \leftrightarrow \bZ_2$,
and $\bZ_{N_c}^\text{1-form}$ corresponds to the center-flavor background $\Omega_{A,B}$ described above. The anomaly \eqref{eq:keyanomaly} excludes the possibility that the phase transition is simply described by \eqref{eq:LG} below the deconfinement temperature
at which the $\bZ_2$ symmetry is broken. 

Next we give a speculative discussions on the case of $0 \leq \mu_B < \pi $. 
Even though there is no $\bZ_2$ symmetry, we may still define confinement in the following technical sense.
Our spacetime is Euclidean, so we can regard $\bR_C$ as a time direction and find a ground state $\ket{\Omega}$ on 
$S^1_T \times S^1_A \times S^1_B$. (This idea is familiar in 2d CFT.)
The boundary conditions $m_{\pm \infty}$ also define physical states $\ket{ m_{\pm \infty} }$.\footnote{
For fermions, we need to impose APS boundary conditions for the APS index theorem to work. They have natural 
Hilbert space interpretations \cite{Yonekura:2016wuc}.}
We define confinement as the statement that $\bra{m=1} \et{\Omega} \neq 0$ as well as $\bra{m=0} \et{\Omega} \neq 0$.
The  $\ket{m=0}$ is expected to always have overlap with $\ket{\Omega}$ for $|\mu_B| \leq \pi $.
In the presence of $\bZ_2$ we have $\bra{m=1} \et{\Omega} = \bra{m=0} \et{\Omega} $ if $\bZ_2$ is unbroken in $\ket{\Omega}$.
So this condition $\bra{m=1} \et{\Omega} \neq 0$ is a generalization of the above case of $\mu_B=\pi $ to any value of $\mu_B$.
Deconfinement means that $\bra{m}\et{\Omega}=0$ for $m \neq 0$.
This criterion of (de)confinement might be supported by 
analytic picture of confinement \cite{Yamazaki:2017ulc}.
Intuitively, confinement means (see e.g., \cite{Witten:2000nv}) that the gauge field and in particular $W_T$ is quantum mechanically well-fluctuating.
Then $\ket{\Omega}$ is a superposition of states with all possible values of $W_T$.
Because $m$ is related to the values of $W_T$ (see \eqref{eq:defm}), we expect $\bra{m}\et{\Omega} \neq 0$ for all $m$ in confining phase.
On the other hand, it is localized near $W_T=I_{N_c}$ in deconfining phase. We leave it as a future work to study more details on this criterion.

If the theory confines in the above technical sense, we do not have a domain wall interpolating $x_C=+\infty$ and $x_C=-\infty$.
Thus the anomaly cannot be matched by a domain wall (i.e., fractional instanton) and hence the chiral symmetry must be broken. 
Then the anomaly constraint \eqref{eq:keyanomaly} works in the same way as in the case of $\mu_B=\pi $.
Again, \eqref{eq:LG} is impossible below the deconfinement temperature. 

Let us make another remark about the effects of $\mu_B$. Because the baryon charge of quarks is $1/N_c$, 
the $\mu_B$ is coupled to quarks via $\mu_B/N_c$. Therefore, in large $N_c$ counting, the effect of $\mu_B/N_c$ is a sub-leading effect.
The inequality $T_{\text{deconf}} \leq T_{\text{chiral}}$ is indeed satisfied~\cite{Aharony:2006da} in a holographic QCD model~\cite{Sakai:2004cn}.
For recent numerical studies at $N_c=3$, see e.g., \cite{Bonati:2016pwz,Philipsen:2016hkv}.


\paragraph{Dual magnetic gauge group.}
The anomaly \eqref{eq:keyanomaly} has implications at zero temperature.
Let us consider supersymmetric QCD.
One of the most remarkable phenomena is the appearance of dual magnetic gauge group in Seiberg's description of those theories~\cite{Seiberg:1994pq,Intriligator:1995au}.
The 't~Hooft anomaly of the usual chiral symmetries was important for those results.

We would like to add one more evidence which may give some new insight.  
Suppose that the chiral symmetry is unbroken. 
Now, if the theory contains only scalars and fermions in low energy after confinement, it cannot match the anomaly \eqref{eq:keyanomaly}.
The key fact here is that the anomaly exists even when the flavor background is flat and that the flat background itself does not produce any fermion zero modes.
Therefore, the anomaly \eqref{eq:keyanomaly} detects the existence of dual magnetic gauge fields.

Now let us apply anomaly matching of \eqref{eq:keyanomaly} to supersymmetric QCD.
\begin{itemize}
\item $N_f \leq N_c$: the chiral symmetry is broken.\footnote{Baryonic branch for $N_f=N_c$ is killed by at least one of the boundary conditions, or by $\mu_B$.}
\item  $N_f=N_c+1$: the chiral symmetry is unbroken, and the theory confines.
In this case, we have ${\rm gcd}(N_c,N_c+1)=1$, so the anomaly  \eqref{eq:keyanomaly} vanishes. Thus dual magnetic gauge group need not appear.
\item  $N_c+2 \leq N_f <3N_c$: the chiral symmetry is unbroken and there appears dual magnetic gauge group $\SU( N_f - N_c )$.
This satisfies the anomaly matching of \eqref{eq:keyanomaly} because ${\rm gcd}(N_c,N_f)={\rm gcd}( N_f - N_c  ,N_f)$.\footnote{
Here we need an assumption that the quantities $m_{\pm \infty} \in \bZ_n$ specifying the boundary conditions are also matched under the Seiberg duality.
However, the precise forms of $(V_A,V_B,V_{\pm \infty})$ appearing in the boundary conditions 
need not be matched under the duality. Only the topological data $m_{\pm \infty}$ matters.
}
\end{itemize}

Our anomaly argument does not rely on supersymmetry at all, so it can also constrain magnetic gauge group in non-supersymmetric QCD.
It would be interesting to apply these constraints to ideas such as hidden local symmetry \cite{Bando:1984ej} (see also \cite{Komargodski:2010mc}).\\~\\
{\bf Note added}: On the same day we submitted our paper to arXiv, two closely related papers \cite{Komargodski:2017smk, Cherman:2017tey} appeared. In \cite{Komargodski:2017smk}, the authors obtained the inequality \eqref{eq:ineqadj} for adjoint QCD by using the same anomaly as ours. In \cite{Cherman:2017tey}, the authors introduced the new order-parameter for QCD by using the mixing of the center and flavor symmetry.

\begin{acknowledgments}

The authors would like to thank Y.~Kikukawa T.~Misumi, K.~Mukaida, S.~Shirai, S.~Sugimoto, and T.~T.~Yanagida for helpful discussions.
The work of KY is supported in part by the WPI Research Center Initiative (MEXT, Japan),
and also supported by JSPS KAKENHI Grant-in-Aid (17K14265). The work of HS is partially supported by the Programs for Leading Graduate Schools, MEXT, Japan, via the Leading Graduate Course for Frontiers of Mathematical Sciences and Physics, and also supported by JSPS Research Fellowship for Young Scientists.

\end{acknowledgments}

\bibliography{ref}


\end{document}